\documentclass[12pt,a4paper,english,superscriptaddress,aps]{revtex4}
\usepackage{amssymb}
\usepackage{amsmath} 
\usepackage{slashed}
\usepackage{graphicx}
\usepackage{babel}
\usepackage{hyperref}
\usepackage{color}
\usepackage{comment}
\makeatletter\AtBeginDocument{\let\@elt\relax}\makeatother

\begin{document}

\title{Dynamical Lorentz Symmetry Breaking in a Scale-free Theory of Gravity}

\author{A. C. Lehum}
\email[]{lehum@ufpa.br}
\affiliation{Faculdade de F\'{i}sica, Universidade Federal do Par\'{a}, 66075-110, Bel\'{e}m, Par\'a, Brazil}
	
\author{J. R. Nascimento}
\email[]{jroberto@fisica.ufpb.br}
\affiliation{Departamento de F\'{\i}sica, Universidade Federal da 
	Para\'{\i}ba\\
	Caixa Postal 5008, 58051-970, Jo\~ao Pessoa, Para\'{\i}ba, Brazil}

\author{A. Yu. Petrov}
\email[]{petrov@fisica.ufpb.br}
\affiliation{Departamento de F\'{\i}sica, Universidade Federal da 
	Para\'{\i}ba\\
	Caixa Postal 5008, 58051-970, Jo\~ao Pessoa, Para\'{\i}ba, Brazil}

\author{P. J. Porf\'{i}rio}\email[]{pporfirio@fisica.ufpb.br}
\affiliation{Departamento de F\'{\i}sica, Universidade Federal da Para\'{\i}ba\\
	Caixa Postal 5008, 58051-970, Jo\~ao Pessoa, Para\'{\i}ba, Brazil}

\begin{abstract}
This paper explores the renormalization of scale-free quadratic gravity coupled to the bumblebee field and its potential for dynamically breaking Lorentz symmetry. We conduct one-loop renormalization of the model and calculate the associated renormalization group functions. {Furthermore, we compute the one-loop effective potential for the bumblebee field, demonstrating that it acquires a non-trivial vacuum expectation value (VEV) through radiative corrections—a phenomenon known as the Coleman-Weinberg mechanism. This spontaneous breaking of scale invariance, driven by the non-zero VEV of the bumblebee field, leads to Lorentz symmetry violation. As a result, the non-minimal coupling between the bumblebee and gravitational fields induces a spontaneous generation of an Einstein-Hilbert term via radiative corrections, suggesting a possible link between the Planck scale and Lorentz violation phenomena.}
\end{abstract}

\maketitle

\section{Introduction}

In contrast to the gauge theories governing the electroweak and strong interactions in the Standard Model (SM), quantizing Einstein's general relativity yields a nonrenormalizable quantum field theory \cite{tHooft:1974toh,Deser:1974zzd,Deser:1974cy}. While it remains feasible to incorporate gravity into the quantum framework by considering energies below the Planck scale \cite{Donoghue:1994dn,Burgess:2003jk,Donoghue:2017pgk}, there have been several attempts to explore quadratic gravity as a candidate for a renormalizable theory of quantum gravity \cite{Stelle:1976gc,Tomboulis:1977jk,Odintsov:1991nd,Salvio:2014soa,Salvio:2017qkx,Einhorn:2014gfa}; for reviews, see Refs. \cite{Buchbinder:2017lnd,Donoghue:2021cza}. Additionally, some cosmological implications of quadratic gravity have been examined in Refs. \cite{Myrzakulov:2014hca,Myrzakulov:2016tsz}. 

Despite their renormalizability, quadratic theories of gravity are known to encounter issues such as the potential appearance of ghosts and tachyons. Recent researches ~\cite{Salvio:2014soa,Alvarez-Gaume:2015rwa,Holdom:2015kbf,Anselmi:2018ibi,Salvio:2018crh,Donoghue:2021cza,Buccio:2024hys} indicate that while ghost states may be manageable, tachyonic states are generally considered an irreparable problem. Donoghue and Menezes noted in Ref.~\cite{Donoghue:2021cza} that actually these theories require more discussions since the existence of such issues is not conclusively established in the literature. This suggests that the viability of these theories as realistic physical models remains an open question, meriting continued investigation into their potential as effective descriptions of quantum gravity.



Specifically, in the study conducted in the Ref. \cite{Salvio:2014soa}, the authors investigate the prospect of a fundamental theory of nature devoid of inherent scale, which is achieved just by taking into account quadratic-curvature invariants in their pure gravity sector \cite{Aoki:2021skm,Aoki:2024jhr,Alvarez-Luna:2022hka}, proposing a renormalizable quantum gravity theory characterized by a graviton kinetic term featuring four derivatives. Consequently, the graviton propagator exhibits a momentum space behavior of $1/p^4$ \cite{Buoninfante:2023ryt}. Within this proposition, the authors postulate the potential for the Planck scale to emerge dynamically at the quantum level (see also \cite{Gialamas:2020snr}). In their proposal, this phenomenon arises from a non-minimal interaction between the scalar field and gravity, denoted by $\xi\phi^2R$, wherein the scalar field assumes a non-zero vacuum expectation value $\langle\phi\rangle\neq0$ as a consequence of radiative corrections.

Within studies of gravitational theories, one of the most interesting issues consists in the formulation of an adequate generalization of gravity to the case of the Lorentz symmetry breaking, which, as it is known \cite{KosSam} can be naturally introduced through a spontaneous symmetry breaking (SSB) mechanism in a low-energy limit of some fundamental theory of nature (notably, string theory), with a subsequent study of perturbative issues in such a theory. It is worth mentioning that the SSB mechanism can explain the origin of possible Lorentz-violating (LV) vectors (tensors) corresponding to different minima of some potential. Moreover, in curved spacetimes, the spontaneous Lorentz symmetry violation (LSV) possesses some advantages in comparison with the explicit one, since, besides providing a consistent mechanism of arising, in this case, there is no restriction for LV  vectors (tensors) to be constants as it is usually assumed in the flat space-time. 
	
The most convenient mechanism for the spontaneous LSV is based on the use of the bumblebee model \cite{KosGra} whose action is composed of the Maxwell-like kinetic term and a potential able to develop spontaneous LSV. The resulting theory in a curved space-time, whose Lagrangian is composed of a sum of bumblebee and gravity ones, and perhaps, the terms involving other fields,  is called the bumblebee gravity (for various issues related to this model, including a detailed discussion of degrees of freedom, see also \cite{KosGra2}; an excellent discussion of conceptual problems regarding LSV in gravity, including the bumblebee models, can be found in \cite{KosLi}). Such a theory has been explored in various contexts, such as checking the consistency of some known gravitational solutions, namely, black hole \cite{Bertolami,Casana2017,ourBH1,ourBH2}, cosmology \cite{Capelo,Maluf2021}, G\"{o}del  and G\"{o}del-type ones \cite{ourgodel,gtype} and wormhole ones \cite{Ovgun}. Besides this, it is worth mentioning studies of dispersion relations in a linearized gravity coupled to the bumblebee field \cite{Maluf2014}. Further, a next step has been done in studies of bumblebee gravity, namely, calculations of perturbative corrections in this theory. Such corrections were obtained, within the metric-affine formalism, in papers \cite{ourbumb1,ourbumb2}.

Since the bumblebee field naturally incorporates spontaneous LSV into standard models, extending the analysis conducted in Ref. \cite{Salvio:2014soa,Salvio:2017qkx} to include scale-free operators based on the bumblebee field is a logical step. One significant aspect is investigating whether the Coleman-Weinberg (CW) mechanism \cite{Coleman:1973jx} can occur when the bumblebee field is coupled to agravity. Among these operators is the non-minimal coupling of the type $B^2R$. If CW mechanism occurs, it allows us to establish a connection between the emergence of the Planck scale and a LV effect. In particular, our study focuses on investigating the occurrence of the Coleman-Weinberg mechanism in the bumblebee-agravity model. 

The structure of the paper looks like follows. In Sec. II, we introduce our Lagrangian and discuss the properties of the necessary projection operators. In Sec. III, we calculate the renormalization group functions in the symmetric phase, followed by the computation of the effective potential in Sec. IV. In Sec. V, we consider possible physical scenarios, and in Sec. VI, we provide the final remarks.

Throughout this paper, we use natural units $c=\hbar=1$.

\section{The bumblebee-agravity Lagrangian}\label{sec11}

An essential aspect in investigating Lorentz symmetry violations involves the bumblebee model~\cite{KosGra},
\begin{eqnarray}
S_B=\int d^4x \left\{ -\frac{1}{4}B^{\mu\nu}B_{\mu\nu}-V\left(B^\mu B_\mu\mp b^2 \right)\right\},
\end{eqnarray}
\noindent where $B_{\mu\nu}=(\partial_{\mu}B_{\nu}-\partial_{\nu}B_{\mu})$ and the potential $V(B^{\mu}B_{\mu}\mp b^2)$ is selected to induce a non-zero VEV for the bumblebee field. This introduces a preferred direction in spacetime, resulting in spontaneous Lorentz symmetry breaking. Typically, the potential takes the form $V=\lambda(B^\mu B_\mu\mp b^2)^2$, where the $\mp$ sign accommodates both space-like and time-like $B^\mu$, while $b^2>0$. 

As said before, the potential is responsible for boosting the Lorentz symmetry breaking what one can be picked to possess a minimum when $b^{\mu}b_{\mu}=\pm b^2$, with $\langle B_{\mu}\rangle=b_{\mu}$. However, as $b_{\mu}$ is a dimensional parameter, the resulting theory naturally introduces a fundamental scale. Our aim in this work is to investigate a theory that avoids fundamental scales from its inception, such that scales could arise from radiative fluctuations via the Coleman-Weinberg mechanism \cite{Coleman:1973jx}.  Keeping this in mind, let us start by taking $b^2=0$  at the classical level to have a scale-independent theory. As a toy model, we propose the couplings of the bumblebee field to the agravity model which can be treated as natural generalizations of couplings defined in \cite{Salvio:2014soa,Salvio:2017qkx}. This theoretical framework is expressed through the action
\begin{eqnarray}\label{eq01}
	\mathcal{S} =&&\!\!\!\!\!\! \int{d^4x}\sqrt{-g}\Big\{ \frac{R^2}{6f_0^2}+\frac{1}{f_2^2}\left(\frac{1}{3}R^2-R^{\mu\nu}R_{\mu\nu}\right) 
	-\frac{1}{4 g_t} B^{\mu\nu} B_{\mu\nu} -\frac{1}{2g_l}  \left(\nabla^\mu B_\mu \right)^2 \nonumber\\
	&+& \xi_1 \left(B^\mu B^\nu - \frac{g^{\mu\nu}}{4}B^2 \right) R_{\mu\nu} + \xi_2 B^2 R - \lambda (B^\mu B_\mu)^2 +\mathcal{L}_{GF} + \mathcal{L}_{FP} + \mathcal{L}_{CT} \Big\},
\end{eqnarray}
\noindent where $R$ denotes the Ricci scalar, $R_{\mu\nu}$ represents the Ricci tensor, and $\nabla^\mu$ stands for the covariant derivative. The constants $\xi_1$ and $\xi_2$ are the couplings associated with the traceless and trace parts of the coupling between the bumblebee field and the gravitational field, respectively. Additionally, $\mathcal{L}_{GF}$ and $\mathcal{L}_{FP}$ denote the gauge-fixing and the corresponding Faddeev-Popov Lagrangian of the gravitational sector, respectively. $\mathcal{L}_{CT}$ represents the Lagrangian of counterterms.

Our subsequent steps involve computing the relevant renormalization group functions to determine the one-loop effective potential for the bumblebee field, utilizing the renormalization group function technique \cite{McKeon:1998tr}. To achieve this, first we must expand $g_{\mu\nu}$ around the flat metric, $g_{\mu\nu}=\eta_{\mu\nu}+ h_{\mu\nu}$, allowing us to express the tree-level Lagrangian, before doing the gauge fixing as
\begin{eqnarray}
\mathcal{L}=\mathcal{L}_h+\mathcal{L}_b+\mathcal{L}_{nm}+\mathcal{L}_{V},
\end{eqnarray}
\noindent where $\mathcal{L}_h$ denotes the quadratic kinetic term of the gravitational Lagrangian
\begin{eqnarray}\label{Lg_quadratic}
	\mathcal{L}_h &=& 
	\frac{1}{8 f_2^2}	\Big[ 
	(\partial^\sigma\partial^\mu-\Box \eta^{\mu\sigma})h_{\mu\nu} (\partial^\rho\partial^\nu-\Box \eta^{\nu\rho})h_{\rho\sigma}
	+(\partial^\rho\partial^\mu-\Box \eta^{\mu\rho})h_{\mu\nu} (\partial^\sigma\partial^\nu-\Box \eta^{\nu\sigma})h_{\rho\sigma}\nonumber\\
	&& 
	-\frac{2}{3}(\partial^\nu\partial^\mu-\Box \eta^{\mu\nu})h_{\mu\nu} (\partial^\rho\partial^\sigma-\Box \eta^{\sigma\rho})h_{\rho\sigma}\Big]
	+\frac{1}{18f_0^2} \left( \partial^{\mu}\partial^\nu h_{\mu\nu}-\Box h \right)^2 + \mathcal{O}(h^3), 
\end{eqnarray}
\noindent $\mathcal{L}_b$ encompasses terms associated with the quadratic portion of the bumblebee field and couplings $g_t$ and $g_l$
\begin{eqnarray}
	\mathcal{L}_b &=& -\frac{1}{4 g_t} B^{\mu\nu}B_{\mu\nu}
	-\frac{1}{2g_l}\partial^\mu B_\mu \partial^\nu B_\nu +\frac{1}{2g_t}B^{\mu\nu}{B^{\tau}}_{\mu}\left(h_{\nu\tau}
	-\frac{1}{4}\eta_{\nu\tau}h \right)\nonumber\\
	&&-\frac{1}{4g_l}\left( \partial^{\mu}B_\mu \partial^{\nu}B_\nu h +2B^\mu \partial^\nu B_\nu \partial_\mu h - 4 B^\mu \partial^\nu B_\nu \partial^\alpha h_{\mu\alpha}- 4 \partial^\mu B_\mu \partial^\nu B^\alpha h_{\nu\alpha} \right)\nonumber\\
	&&-\frac{1}{32g_t}\left(h^2-2h^{\mu\nu}h_{\mu\nu} \right) B^{\alpha\beta}B_{\alpha\beta}
	-\frac{1}{2g_l} B^\mu B^\nu \left(\partial^\alpha h_{\mu\alpha} \partial^\beta h_{\nu\beta} -\partial^\alpha h_{\mu\alpha} \partial^\nu h
	+\frac{1}{4} \partial_\mu h \partial_\nu h \right)\nonumber\\
	&&-\frac{1}{2g_l}B^\mu \partial^\beta B_{\beta}\left(2 \partial_\nu h^{\alpha\nu}h_{\mu\alpha} - \partial^\nu h h_{\mu\nu} 
	- \partial^\nu h_{\mu\nu} h +\frac{1}{2} \partial_\mu h h
	+2 \partial_\alpha h_{\mu\nu} h^{\nu\alpha} - \partial_\mu h^{\alpha\nu} h_{\alpha\nu}\right)\nonumber\\
	&&-\frac{1}{2g_l} \partial^\alpha B_{\alpha} \left( 2 \partial^{\mu}B_\nu h^{\nu\beta} h_{\mu\beta} -\partial^{\mu}B_\nu h^{\mu\nu} h
	-\frac{1}{4} \partial^{\mu}B_\mu h^{\nu\beta} h_{\nu\beta}
	+\frac{1}{8} \partial^{\mu}B_\mu h^2  
	\right), \nonumber\\
	&&-\frac{1}{2g_l}B^\mu \partial^\alpha B^\nu\left(2\partial^\beta h_{\mu\beta} h_{\nu\alpha} - \partial_\mu h h_{\nu\alpha} \right)
	-\frac{1}{2g_l}\partial^\mu B^\nu \partial^\alpha B^\beta h_{\mu\nu} h_{\alpha\beta}+\mathcal{O}(h^3), 
\end{eqnarray} 
\noindent $\mathcal{L}_{nm}$ comprises terms arising from couplings $\xi_1$ and $\xi_2$
\begin{eqnarray}
	\mathcal{L}_{nm} &=& \xi_1 B^{\alpha}B^{\beta}\left( 
	\partial^{\gamma}\partial_{\beta}h_{\alpha\gamma}-\partial_{\alpha}\partial_{\beta}h -\frac{1}{2}\Box h_{\alpha\beta}\right) 
	+ \left(\xi_2-\frac{\xi_1}{4}\right) B^2 \left(
	\partial^{\gamma}\partial^{\beta}h_{\gamma\beta}-\Box h		
	\right)\nonumber\\
	&&+\xi_1 B^\alpha B^\beta \Big(\frac{1}{2}\partial^\nu {h_{\alpha}}^\mu \partial_\nu h_{\beta\mu} 
	-\frac{1}{2}\partial^\nu h_{\alpha\mu} \partial^\mu h_{\beta\nu}
	+\frac{1}{4}\partial^\alpha h^{\mu\nu} \partial_\beta h_{\mu\nu}
	+\frac{1}{2}\partial^\nu h_{\alpha\beta} \partial^\mu h_{\nu\mu}\nonumber\\
	&& - \partial_\beta {h_{\alpha}}^{\mu} \partial^\nu h_{\mu\nu}
	-\frac{1}{4}\partial^\nu h_{\alpha\beta} \partial_\nu h
	+\frac{1}{2}\partial_\beta h_{\alpha\mu} \partial^\mu h
	+\Box {h_{\beta}}^{\mu} h_{\alpha\mu}
	-\partial^\nu \partial^\mu h_{\beta\mu} h_{\alpha\nu}
	-\frac{1}{4}\Box h_{\alpha\beta} h\nonumber\\
	&& -\partial_\beta \partial^\nu {h_{\nu}}^{\mu} h_{\alpha\mu}
	+\partial^\mu \partial_\beta h h_{\alpha\mu} 
	+\frac{1}{2}\partial^\nu \partial_\beta h_{\alpha\nu}h
	-\frac{1}{4}\partial_\alpha \partial_\beta h h
	+\frac{1}{2}\partial^\nu \partial^\mu h_{\alpha\beta}h_{\mu\nu}\nonumber\\
	&&	-\partial^\nu \partial_\beta {h_{\alpha}}^{\mu}h_{\mu\nu}
	+\frac{1}{2}\partial_\beta \partial^\alpha h^{\mu\nu}h_{\mu\nu}
	\Big)
	+\left(\xi_2-\frac{\xi_1}{4}\right) B^{\alpha}B^{\beta}\left(\Box h h_{\alpha\beta}-\partial^\mu \partial^\nu h_{\mu\nu} h_{\alpha\beta}\right) \nonumber\\
	&& 
	+\left(\xi_2-\frac{\xi_1}{4}\right) B^2 \Big( \frac{3}{4} \partial^\alpha h^{\beta\mu} \partial_\alpha h_{\beta\mu}
	-\frac{1}{2} \partial^\alpha h^{\beta\mu} \partial_\mu h_{\beta\alpha}
	+\partial^\alpha h \partial^\mu h_{\mu\alpha}
	- \partial^\alpha h^{\alpha\beta} \partial^\mu h_{\beta\mu}\nonumber\\
	&&
	-\frac{1}{2}  \Box h h
	 -\frac{1}{4}  \partial^\alpha h \partial_\alpha h 
	+\frac{1}{2}  \partial^\alpha \partial^\beta h_{\alpha\beta} h
	+\Box h^{\mu\nu} h_{\mu\nu} 
	-2 \partial^\alpha \partial_\beta h_{\alpha\mu} h^{\mu\beta}
	+\partial^\alpha \partial^\beta h h_{\alpha\beta}
	\Big)\nonumber\\
	&&+\mathcal{O}(h^3),
\end{eqnarray}
\noindent and $\mathcal{L}_V$ stands for the bumblebee potential and its gravitational interactions
\begin{eqnarray}
	\mathcal{L}_{V} &=& \lambda B^{\alpha}B^{\beta} B^{\mu}B^{\nu} \Big(
	\eta_{\alpha\beta}\eta_{\mu\nu}
	+\frac{1}{2}\eta_{\alpha\beta}\eta_{\mu\nu}h
	-2 \eta_{\alpha\beta}h_{\mu\nu}
	+h_{\alpha\beta}h_{\mu\nu}
	+2\eta_{\alpha\beta} {h_{\mu}}^{\gamma}h_{\nu\gamma}\nonumber\\
	&&
	-\frac{1}{4}\eta_{\alpha\beta}\eta_{\mu\nu} h^{\gamma\tau} h_{\gamma\tau} 
	-\eta_{\alpha\beta} h_{\mu\nu}h+\frac{1}{4}\eta_{\alpha\beta}\eta_{\mu\nu}h^2\Big)
	+\mathcal{O}(h^3).
\end{eqnarray}
In a certain sense, our study can be treated as an analogue of that one performed in \cite{KosGra3}, where the weak field limit was studied in the standard Einstein-bumblebee gravity.

To quantize the model, we introduce the gauge-fixing Lagrangian given by 
\begin{equation}
\mathcal{L}_{GF}=-\frac{1}{2\zeta_g}\partial^\nu( h_{\mu\nu}-\frac{1}{2}\eta_{\mu\nu}h)\partial_\alpha( h^{\mu\alpha}-\frac{1}{2}\eta^{\mu\alpha}h).
\end{equation}
\noindent Consequently, the quadratic part of the action leads to the following propagators:
\begin{eqnarray}
\Delta^{\mu\nu}(p) &=& -\frac{i}{p^2} \left[ g_t T^{\mu\nu} + g_l L^{\mu\nu} \right]; \nonumber\\
\Delta_{\mu\nu\rho\sigma}(p) &=& \frac{i}{p^4} \left[ -2f_2^2 P^{(2)}_{\mu\nu\rho\sigma} + f_0^2 P^{(0)}_{\mu\nu\rho\sigma} + 2\zeta_g \left( P^{(1)}_{\mu\nu\rho\sigma} + \frac{1}{2} P^{(0w)}_{\mu\nu\rho\sigma} \right) \right],
\end{eqnarray}
where $\Delta^{\mu\nu}(p)$  and $\Delta_{\mu\nu\rho\sigma}(p)$ are the bumblebee and graviton propagators, respectively. The projectors are defined as follows:
\begin{eqnarray}
P^{(2)}_{\mu\nu\rho\sigma} &=& \frac{1}{2} T_{\mu\rho} T_{\nu\sigma} + \frac{1}{2} T_{\mu\sigma} T_{\nu\rho} - \frac{1}{D-1} T_{\mu\nu} T_{\sigma\rho}; \nonumber\\
P^{(1)}_{\mu\nu\rho\sigma} &=& \frac{1}{2} \left( T_{\mu\rho} L_{\nu\sigma} + T_{\mu\sigma} L_{\nu\rho} + L_{\mu\rho} T_{\nu\sigma} + L_{\mu\sigma} T_{\nu\rho} \right); \nonumber\\
P^{(0)}_{\mu\nu\rho\sigma} &=& \frac{1}{D-1} T_{\mu\nu} T_{\sigma\rho}; \nonumber\\
P^{(0w)}_{\mu\nu\rho\sigma} &=& L_{\mu\nu} L_{\sigma\rho},
\end{eqnarray}
where $T_{\mu\nu}$ and $L_{\mu\nu}$ are given by
\begin{eqnarray}
T_{\mu\nu} &=& \eta_{\mu\nu} - \frac{p_\mu p_\nu}{p^2}; \nonumber\\
L_{\mu\nu} &=& \frac{p_\mu p_\nu}{p^2}.
\end{eqnarray}
\noindent These projectors will be further employed in our calculations. 

In this work, we treat the model using the one-graviton exchange approximation, which involves considering the lowest non-trivial gravitational correction. In this scenario, the ghost action decouples completely, rendering its contribution to the quantum corrections trivial. For this reason, we omit the graviton ghost Lagrangian $\mathcal{L}_{FP}$ from our analysis.

\section{Renormalization group functions}

\begin{figure}[ht!]
	\includegraphics[angle=0 ,width=8cm]{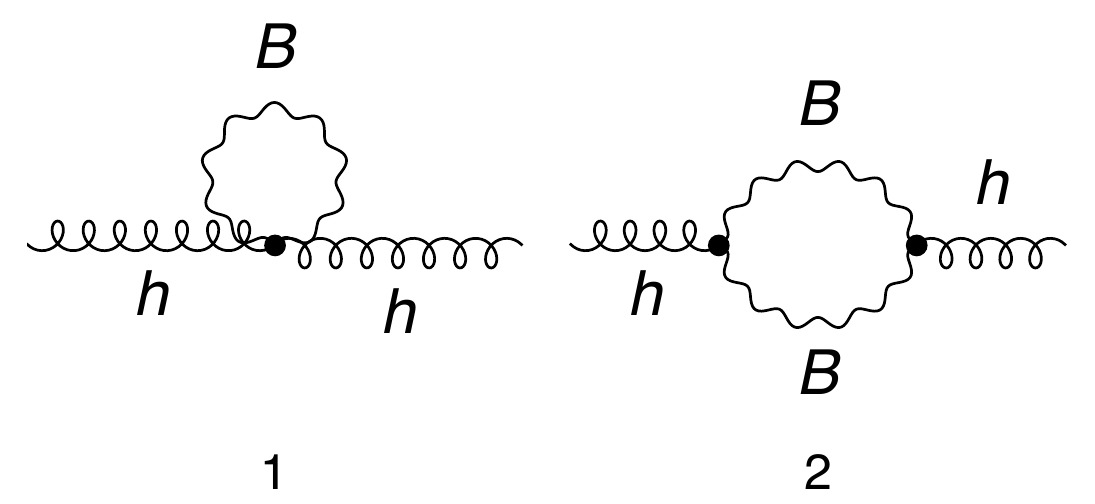}
	\caption{Bumblebee corrections to the graviton propagation. Wavy and wiggly lines represent the bumblebee and graviton propagators, respectively.}
	\label{fig01}
\end{figure}

In this section, we present the UV renormalization of the model. We begin with the bumblebee corrections to the graviton propagation. The Feynman diagrams are illustrated in Figure \ref{fig01}. In order to compute the Feynman diagrams we used a set of Mathematica\textsuperscript{TM} packages \cite{feyncalc,feyncalc1,feyncalc2,feynarts,feynrules,feynrules1,feynhelpers}. 

It is evident that diagram \ref{fig01}.1 vanishes since the $B$ field is massless. Upon evaluating diagram \ref{fig01}.2, we obtain a tensorial structure identical to that derived from the gravitational kinetic term presented in \eqref{Lg_quadratic}, which can thus be interpreted as a renormalization of the gravitational couplings $f_0$ and $f_2$. The expression for the UV divergent part of diagram \ref{fig01}.2 is given by
\begin{eqnarray}\label{graviton_SE}
i\Gamma_{\mu\nu\alpha\gamma}(p) &=& 
\frac{1}{6} P^{(2)}_{\mu\nu\alpha\gamma}
\Big[\frac{\delta_{f_2^2}}{f_2^2} -
\frac{1}{960 \pi^2 \epsilon }\Big(5\xi_1^2 \left(g_l^2+4 g_l g_t+7 g_t^2\right)-10\xi_1 \left(g_l+5 g_t\right)+7\Big)\Big] \nonumber\\
&& +\frac{1}{9} P^{(0)}_{\mu\nu\alpha\gamma} \Big[\frac{\delta_{f_0^2}}{f_0^2} +
\frac{1}{128 \pi^2 \epsilon }\Big(\xi_1^2 \left(g_l^2+4 g_l g_t+7 g_t^2\right)+144 \xi_2^2 \left(g_l^2+3 g_t^2\right)\nonumber\\
&&-2 \xi_1 (g_l +5g_t)+24 \xi_2(g_l-3 g_t)+2\Big)\Big], 
\end{eqnarray}
from which, by applying the MS scheme of renormalization to impose finiteness, we can easily derive the counterterms $\delta_{f_0^2}$ and $\delta_{f_2^2}$. Notably, the bumblebee loops maintain the transverse propagation of the graviton.

\begin{figure}[ht!]
	\includegraphics[angle=0 ,width=12cm]{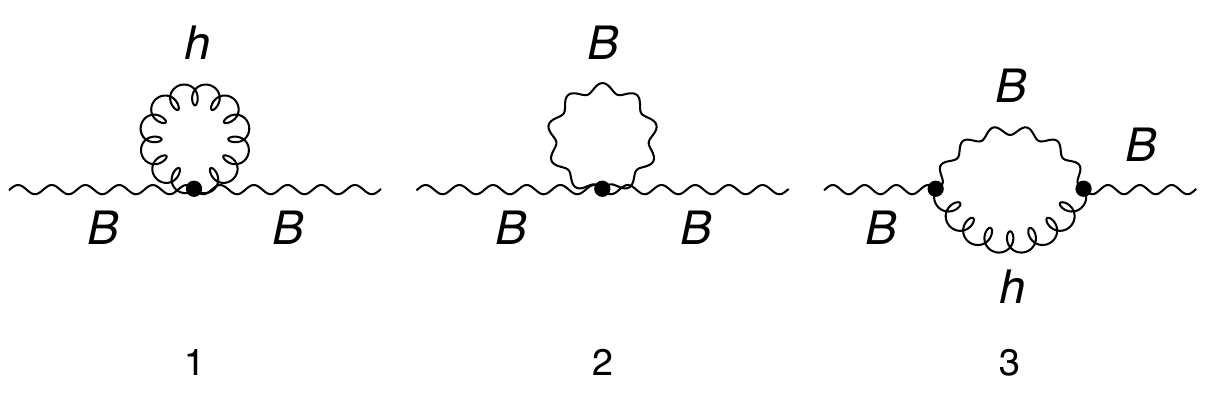}
	\caption{Bumblebee self-energy.}
	\label{fig02}
\end{figure}

Our next step involves computing the bumblebee field self-energy. The corresponding Feynman diagrams are depicted in Figure \ref{fig02}. The UV divergent contribution is given by
\begin{eqnarray}\label{bumblebee_SE}
i\Gamma_{\mu\nu}(p) &=& -p^2\Big[ T_{\mu\nu}\left( \frac{\delta_{g_t}}{g_t} - \Gamma_{T}\right)
-L_{\mu\nu} \left( \frac{\delta_{g_l} }{g_l}-\Gamma_L\right) \Big],	
\end{eqnarray}

\noindent where
\begin{eqnarray}
\Gamma_T &=&	\frac{g_t}{288 \pi^2 \epsilon } 
	 \Big[ \xi_1^2\left(f_0^2 (5 g_l-4 g_t)+5 f_2^2 (2 g_l-g_t)\right) +4 \xi_1 \left(f_0^2 (6 g_l \xi_2-3 g_t \xi_2-2)+5 f_2^2\right)\nonumber\\ &+&
	 48 f_0^2 \xi_2
	\Big];\nonumber\\
\Gamma_L &=& \frac{1}{96\pi^2 g_l\epsilon} \Big[
f_0^2 (\xi_1 (12 g_l^2 \xi_2-24 g_l g_t \xi_2+5g_l-3g_t) + g_l \xi_1^2 (g_t -2g_l)-12g_l \xi_2 +36g_t\xi_2-3)\nonumber\\
&& +5f_2^2(g_l\xi_1^2(g_l-2g_t)+2\xi_1(g_l+3g_t)-3)
 \Big].
\end{eqnarray} 
By imposing finiteness through the MS scheme, we find the counterterms to be $\delta_{g_t}=g_t\Gamma_T$ and $\delta_{g_l}=g_l\Gamma_L$. 

It is important to note that the coupling constants $g_t$ and $g_l$, similar to $f_0$ and $f_2$ in the context of graviton self-energy, are responsible for renormalizing the self-energy of the bumblebee field, thereby eliminating the need for wave-function renormalization of the bumblebee field.

\begin{figure}[ht!]
	\includegraphics[angle=0 ,width=8cm]{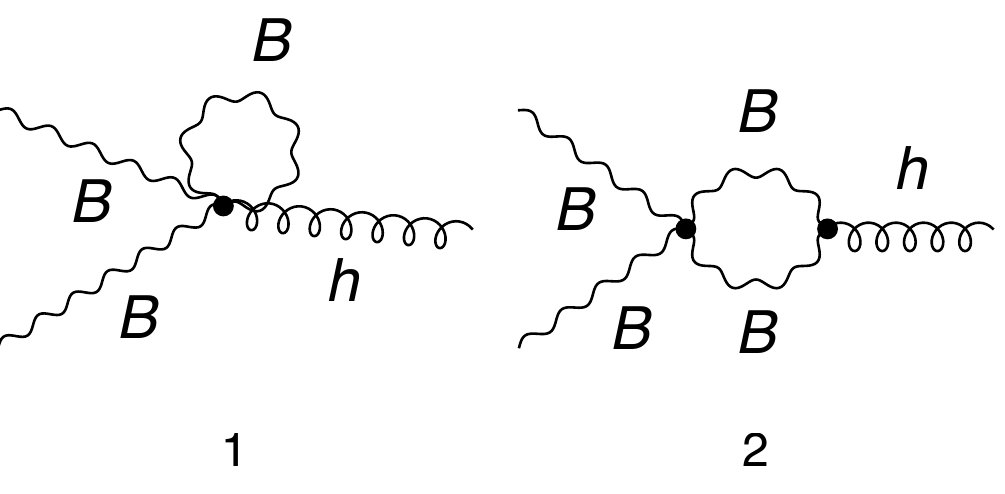}
	\caption{Three-point function. This function gives the renormalization factor of $\xi_1$ and $\xi_2$.}
	\label{fig03}
\end{figure}

In the following, we compute the renormalization of the non-minimal couplings $\xi_1$ and $\xi_2$. The necessary diagrams to compute the renormalization factors of $\xi_1$ and $\xi_2$ are depicted in Figure \ref{fig03}. The corresponding UV divergent part of the three-point function $\Gamma^{\alpha\gamma\mu\nu}(p_1,p_2,p_3)=\langle T B^{\alpha}(p_1)B^{\gamma}(p_2)h^{\mu\nu}(p_3) \rangle$ is given by
\begin{eqnarray}
-i\Gamma^{\alpha\gamma\mu\nu}(p_1,p_2,p_3)&=&\frac{1}{2}\left[p_3^2 (\eta^{\alpha  \mu } \eta^{\gamma  \nu }-\eta^{\alpha  \gamma } \eta^{\mu  \nu })
+\eta^{\gamma  \mu } \left( p_3^2 \eta^{\alpha  \nu }-p_3^{\alpha } p_3^{\nu }\right)+p_3^{\gamma } \left(2 p_3^{\alpha } \eta^{\mu  \nu }-p_3^{\nu } \eta^{\alpha  \mu }\right)\right.+\nonumber\\ &+& \left.
p_3^{\mu } \left( p_3^{\nu} \eta^{\alpha\gamma}-p_3^{\alpha } \eta^{\gamma  \nu }
-p_3^{\gamma } \eta^{\alpha  \nu }\right)\right] \Big[ \delta_{\xi_1} - \frac{\lambda  \left(
\xi_1 (g_l^2+4g_l g_t+7g_t^2)-g_l-5g_t\right)}{24 \pi^2 \epsilon}
\Big]\nonumber\\
&+&
2 \eta^{\alpha \gamma} \left(p_3^2 \eta^{\mu \nu }-p_3^{\mu} p_3^{\nu }\right)  \Big[\delta_{\xi_2}-\frac{\lambda \left(12\xi_2 (g_l^2+3g_t^2) +g_l-3g_t\right)}{24 \pi^2 \epsilon} \Big]+\mathrm{finite}.
\end{eqnarray}

Thus, using the MS scheme, the renormalization factors $\delta_{\xi_1}$ and $\delta_{\xi_2}$ can be readily calculated. Specifically, for $\xi_1 = \xi_2 = 0$, the renormalization conditions indicate that the counterterms $\delta_{\xi_1}$ and $\delta_{\xi_2}$ assume particular forms. This underscores the critical role of non-minimal couplings from the outset in ensuring the renormalizability of the model.

\begin{figure}[ht!]
	\includegraphics[angle=0 ,width=16cm]{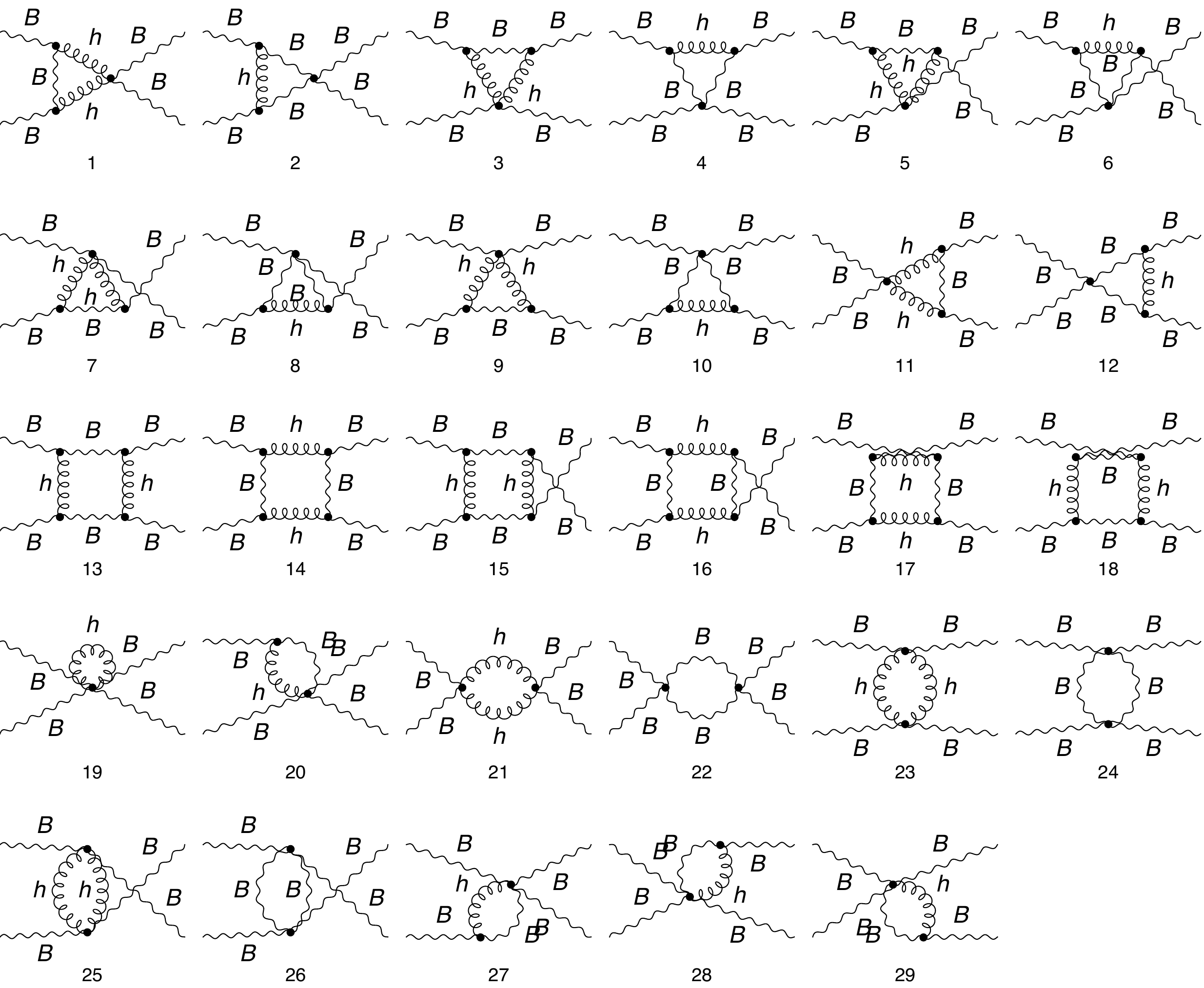}
	\caption{The one-loop bumblebee four-point function up to order $\xi_i^2$. This function gives the renormalization of the bumblebee self-coupling $\lambda$.}
	\label{fig04}
\end{figure}

Finally, we now proceed to calculate the renormalization factor for the four-point self-coupling function of the bumblebee field. This calculation is conducted up to quadratic order in the non-minimal couplings $\xi_1$ and $\xi_2$. The one-loop bumblebee four-point function is illustrated in Figure \ref{fig04}. The associated UV divergent part is expressed as
\begin{eqnarray}
\langle T B^{\mu}(p_1)B^{\nu}(p_2)B^{\alpha}(p_3)B^{\gamma}(p_4) \rangle =
(\eta^{\nu\alpha}\eta^{\mu\gamma}
+\eta^{\mu\alpha}\eta^{\nu\gamma} 
+\eta^{\nu\mu}\eta^{\alpha\gamma})i\Gamma^{(4)}(p_1,p_2,p_3,p_4),
\end{eqnarray}
\noindent where  
 \begin{eqnarray}
\Gamma^{(4)}(p_1,p_2,p_3,p_4) &=& \frac{\lambda^2(5g_l^2+2 g_l g_t+17 g_t^2)}{\pi^2\epsilon}+\frac{\lambda(f_0^2-10f_2^2)}{6\pi^2\epsilon}+\frac{\lambda\xi_1(f_0^2(g_t+3g_l)-40 f_2^2 g_t)}{3\pi^2\epsilon} \nonumber\\
&&-\frac{2\lambda\xi_2 f_0^2(g_l-g_t)}{\pi^2\epsilon}
+\frac{\lambda\xi_1^2(f_0^2(9g_l^2+3g_l g_t +4g_t^2)-100f_2^2 g_t^2)}{24\pi^2\epsilon}\nonumber\\
&&+\frac{6\lambda\xi_2^2f_0^2(g_l^2+g_l g_t+4g_t^2)}{\pi^2\epsilon}
+\frac{\lambda\xi_1\xi_2 f_0^2 (3g_l^2+g_l g_t-4 g_t^2)}{\pi^2 \epsilon}\nonumber\\
&& +\frac{\xi_1^2(13f_0^4-5f_0^2 f_2^2+325 f_2^4)}{1152\pi^2\epsilon}
+\frac{5\xi_2^2(f_0^4-f_0^2 f_2^2+f_2^4)}{8\pi^2 \epsilon} \nonumber\\
&& +\frac{\xi_1 \xi_2(8f_0^4-5f_0^2 f_2^2+30 f_2^4)}{48\pi^2\epsilon}-8\delta_\lambda,
 \end{eqnarray}
\noindent with the last term arising from the counterterm diagram. It is crucial to emphasize that while there may be gravitational gauge dependence within individual diagrams depicted in Figure \ref{fig04}, the aggregate amplitude becomes gauge-independent upon summing all diagrams. The counterterm $\delta_\lambda$ is determined by imposing the condition of finiteness on the above equation.

With all counterterms evaluated, we can proceed to determine the renormalization group functions. The beta functions for the coupling constants of the model are calculated using the counterterms along with the following relationships between the bare and renormalized couplings:
\begin{subequations}
\begin{eqnarray}
\lambda_0 &=& \mu^{2\epsilon} Z_\lambda \lambda=\mu^{2\epsilon} (\lambda+\delta_\lambda);\\
{\xi_1}_0 &=& \mu^{2\epsilon}Z_{\xi_1}\xi_1=\mu^{2\epsilon}(\xi_1+\delta_{\xi_1});\\
{\xi_2}_0 &=& \mu^{2\epsilon}Z_{\xi_2}\xi_2=\mu^{2\epsilon}(\xi_2+\delta_{\xi_2});\\
\frac{1}{{g_t}_0}&=& \frac{\mu^{2\epsilon} Z_{g_t}}{g_t}=\frac{\mu^{2\epsilon}(1+\delta_{g_t})}{g_t};\\
\frac{1}{{g_l}_0}&=& \frac{\mu^{2\epsilon} Z_{g_l}}{g_l}=\frac{\mu^{2\epsilon}(1+\delta_{g_l})}{g_l};\\
\frac{1}{{f_0}_0^2}&=& \frac{\mu^{2\epsilon}(1+\delta_{f_0^2})}{f_0^2};\\
\frac{1}{{f_2}_0^2}&=&\frac{\mu^{2\epsilon}(1+\delta_{f_2^2})}{f_2^2}.
\end{eqnarray}
\end{subequations}

These relationships, together with the expressions for the counterterms, lead to the following beta functions:
\begin{subequations}
\begin{eqnarray}
	\beta(\lambda)&=&\frac{\lambda^2(5g_l^2+2 g_l g_t+17 g_t^2)}{4\pi^2}+\frac{\lambda(f_0^2-10f_2^2)}{24\pi^2}+\frac{\lambda\xi_1(f_0^2(g_t+3g_l)-40 f_2^2 g_t)}{12\pi^2} \nonumber\\
	&&-\frac{\lambda\xi_2 f_0^2(g_l-g_t)}{2\pi^2}
	+\frac{\lambda\xi_1^2(f_0^2(9g_l^2+3g_l g_t +4g_t^2)-100f_2^2 g_t^2)}{96\pi^2}\nonumber\\
	&&+\frac{3\lambda\xi_2^2f_0^2(g_l^2+g_l g_t+4g_t^2)}{2\pi^2}
	+\frac{\lambda\xi_1\xi_2 f_0^2 (3g_l^2+g_l g_t-4 g_t^2)}{4\pi^2}\nonumber\\
	&& +\frac{\xi_1^2(13f_0^4-5f_0^2 f_2^2+325 f_2^4)}{4608\pi^2}
	+\frac{5\xi_2^2(f_0^4-f_0^2 f_2^2+f_2^4)}{32\pi^2} \nonumber\\
	&& +\frac{\xi_1 \xi_2(8f_0^4-5f_0^2 f_2^2+30 f_2^4)}{192\pi^2};\label{betalambda}\\
 \beta(\xi_1) &=& \frac{\lambda  \left(
		\xi_1 (g_l^2+4g_l g_t+7g_t^2)-g_l-5g_t\right)}{12 \pi^2} ;\label{betaxi1}\\
	\beta(\xi_2)&=&\frac{\lambda \left(12\xi_2 (g_l^2+3g_t^2) +g_l-3g_t\right)}{12 \pi^2};\label{betaxi2}\\
 \beta(g_t) &=&	\frac{g_t^2}{144 \pi^2} 
	\Big[ \xi_1^2\left(f_0^2 (5 g_l-4 g_t)+5 f_2^2 (2 g_l-g_t)\right) \nonumber\\ &&+ 4 \xi_1 \left(f_0^2 (6 g_l \xi_2-3 g_t \xi_2-2)+5 f_2^2\right)+
	48 f_0^2 \xi_2
	\Big];\label{betagt} \\
	\beta(g_l) &=& \frac{g_l}{48\pi^2} \Big[
	5f_2^2(g_l\xi_1^2(g_l-2g_t)+2\xi_1(g_l+3g_t)-3)\nonumber\\
	&&+
	f_0^2 \Big( \xi_1 (12 g_l^2 \xi_2-24 g_l g_t \xi_2+5g_l-3g_t) \nonumber\\
	&&+ g_l \xi_1^2 (g_t -2g_l) -12g_l \xi_2 +36g_t\xi_2-3 \Big) 
	\Big];\label{betagl} \\
 \beta(f_0^2) &=&	
	 \frac{5}{96\pi^2}(f_0^4+6f_0^2 f_2^2+10f_2^4)-
	 \frac{f_0^2}{64 \pi^2}
	 \Big[\xi_1^2 \left(g_l^2+4 g_l g_t+7 g_t^2\right)\nonumber\\
	 &&+144 \xi_2^2 \left(g_l^2+3 g_t^2\right)-2 \xi_1 (g_l +5g_t)+24 \xi_2(g_l-3 g_t)+2\Big];\label{betaf0}\\
	\beta(f_2^2)
	&=&
	-\frac{133f_2^4}{160\pi^2}+\frac{f_2^2}{480 \pi^2}\Big[5\xi_1^2 \left(g_l^2+4 g_l g_t+7 g_t^2\right)-10\xi_1 \left(g_l^2+5 g_t\right)+7\Big],\label{betaf2}
\end{eqnarray}
\end{subequations}
\noindent where we have borrowed the first term on the {\textit{r.h.s.}} of Eqs. (\ref{betaf0}) and (\ref{betaf2}) from Ref. \cite{Salvio:2014soa}, which corresponds to contributions from graviton self-interactions and gravitational ghost fluctuations. It is noteworthy that even if $\lambda = 0$ at tree level, $\beta(\lambda)$ remains nontrivial. This behavior is analogous to the beta function of the self-coupling scalar field in Scalar QED \cite{Srednicki:2007qs}, where it is proportional to $e^4$ when $\lambda = 0$ at tree level.

In our analysis, we adopt the interpretation from Ref. \cite{Salvio:2014soa}, where the UV divergences of the one-loop graviton self-energy are absorbed by the renormalization of the gravitational couplings $f_0$ and $f_2$. Similarly, we extend this approach to the bumblebee self-energy by absorbing its UV divergences through the redefinition of the couplings $g_t$ and $g_l$. Consequently, we find that the anomalous dimensions for both the graviton and bumblebee fields vanish, at least in one-loop order, thereby ensuring consistency in our treatment of renormalization across these two fields.

In the next section, we will use these beta functions to compute the effective potential and explore the possibility of emergence of a LV phase.

\section{The broken Lorentz symmetry phase: Coleman-Weinberg mechanism}

Given our examination of the UV behavior of the bumblebee field field coupled to agravity, we can now endeavor to comprehend the potential for dynamical LSB through the CW mechanism~\cite{Coleman:1973jx}.

To do this, we will calculate the one-loop effective bumblebee potential using the renormalization group method \cite{McKeon:1998tr}. This method offers a robust approach for computing the improved leading-log effective potential and has been extensively employed across various contexts, as evidenced by its widespread utilization in the literature \cite{Elias:2003zm,Elias:2004bc,Dias:2010it,Lehum:2019msl,Souza:2020hjd,Lehum:2023qnu}. It is noteworthy that there exists a discrepancy between the renormalization group functions evaluated in the CW scheme and those in the MS scheme~\cite{Chishtie:2007vd}, though this discrepancy is not relevant to this study since our focus is solely on computing the effective potential up to one-loop order. Additionally, in this approximation, we need not concern ourselves with the emergence of gauge-dependent objects such as daisies \cite{Bazeia:1988pz,deLima:1989yf}.

Driven by dimensional analysis, the perturbative expansion of the effective potential exhibits the general form
\begin{eqnarray}\label{veff_pert}
V_{\text{eff}}(B_c^2)=A_0(x) B_c^4 + A_1(x) B_c^4 L+ A_{2}(x) B_c^4 L^2+\cdots,
\end{eqnarray}
\noindent where $x$ represents the collection of coupling constants, $L=\ln\left(B_c^2/\mu^2\right)$, $B_c^\mu$ is the classical bumblebee field and $\mu$ denotes an energy scale introduced by the regularization procedure. The coefficients $A_i=(a_0^{(i)}x+a_1^{(i)}x^2+a_2^{(i)}x^3+\cdots)$ are power series of the coupling constants $x$, computed order by order in the perturbative loop expansion, with the index $i$ denoting a specific loop.

The full effective potential can be reorganized into a leading-logarithm (LL) series as follows:
\begin{eqnarray}\label{ansatz_veff}
	V_{\text{eff}}(B_c^2) = B_c^4 \left( \sum_{n=0}^{\infty} C_{LL}^{(n)}(x) L^n + \sum_{n=0}^{\infty} C_{NLL}^{(n)}(x) L^{n} + \cdots \right),
\end{eqnarray}
where $C_{LL}^{(n)}(x)=a_n^{(n)}x^{n+1}$ and $C_{NLL}^{(n)}(x)=a_{n+1}^{(n)}x^{n+2}$ represent the LL and next-to-leading-logarithm (NLL) coefficients, respectively.

It is important to note that in Eq. \eqref{veff_pert}, the expansion is expressed as a series in which arbitrary powers of $x$ multiply each power of $L$, thereby providing a more general representation of the effective potential. In contrast, Eq. \eqref{ansatz_veff} reorganizes the same series to explicitly display the contributions in terms of LL and NLL terms, which manifest naturally as $x^{n+1} L^n$, $x^{n+2} L^n$, and so on. This reorganization follows standard practices in RGE effective potential calculations, ensuring consistency between the two equations while making the structure of the leading logarithmic contributions clearer.

To utilize the RGE for deriving the effective potential, it is crucial to recognize that $V_{\text{eff}}(B_c^2,x)$ must be independent of the regularization scale $\mu$. Therefore, $V_{\text{eff}}(B^2,x)$ must adhere to the condition
\begin{eqnarray}\label{eq_rge}
\mu\frac{dV_{\text{eff}}}{d\mu}&=&\left(\mu\frac{\partial}{\partial\mu} + \mu\frac{\partial x}{\partial\mu}\frac{\partial}{\partial x}+\mu\frac{\partial B_c^2}{\partial\mu}\frac{\partial}{\partial B_c^2} \right) V_{\text{eff}}\nonumber\\
&=&\left(\mu\frac{\partial}{\partial\mu} + \beta(x)\frac{\partial}{\partial x}+2\gamma B_c^2\frac{\partial}{\partial B_c^2} \right) V_{\text{eff}}=0,
\end{eqnarray}	  
\noindent where $\gamma=\frac{1}{2}\frac{d\ln{Z_3}}{d\ln\mu}$ is the anomalous dimension of the bumblebee field. 

Considering $\mu\frac{\partial V_{\text{eff}}}{\partial\mu}=-2 \frac{\partial V_{\text{eff}}}{\partial L}$ and $B_c^2\frac{\partial V_{eff}}{\partial B_c^2}=2V_{\text{eff}}+\frac{\partial V_{\text{eff}}}{\partial L}$, we rewrite the RGE \eqref{eq_rge} as  
\begin{eqnarray}\label{eq_rge1}
\left(2(\gamma-1)\frac{\partial}{\partial L} + \beta(x)\frac{\partial}{\partial x}+4\gamma\right) V_{\text{eff}}=0.
\end{eqnarray} 

Assuming that the anomalous dimension of the bumblebee field $\gamma$ is vanishing, as discussed in the previous section, we can insert the \emph{ansatz} \eqref{ansatz_veff} into the RGE \eqref{eq_rge1}. In the LL approximation, we find the following recursive relation:
\begin{eqnarray}
C_{LL}^{(n)}(x)=\frac{1}{2n}\beta(x)\frac{\partial C_{LL}^{(n-1)}(x)}{\partial x}, ~~\mathrm{for}~1\le n.
\end{eqnarray} 

To compute the one-loop effective potential, we only need to determine the $C_{LL}^{(1)}(x)$ coefficient. Notably, to reproduce the classical potential $C_{LL}^{(0)}(x)=\lambda$, $C_{LL}^{(1)}(\lambda)$ is linked to the one-loop beta function of $\lambda$, denoted as $\beta^{1l}(\lambda)$, as $\beta^{1l}(\lambda)/2$. This relation allows us to express the one-loop effective potential as follows:
\begin{eqnarray}
	V_{\text{eff}}=(\lambda+\delta)B_c^4+\frac{\beta(\lambda)}{2}B_c^4\ln\left(\frac{B_c^2}{\mu^2} \right),
\end{eqnarray}
where $\delta$ represents a finite counterterm required to satisfy the CW renormalization condition:
\begin{eqnarray}
\frac{d^2V_{0}}{d(B_c^2)^2}=\frac{d^2V_{\text{eff}}}{d(B_c^2)^2}\Big{|}_{B_c^2=v^2}=2\lambda,
\end{eqnarray}
\noindent with $V_0=\lambda B_c^4$ being the classical bumblebee potential and $v$ standing for the renormalization scale.

Thus, the CW renormalized effective potential is given by
\begin{eqnarray}\label{eqVCW}
V_{\text{CW}}= B_c^4\left[\lambda-\frac{3\beta(\lambda)}{4}+\frac{\beta(\lambda)}{2}\ln\left(\frac{B_c^2}{v^2}\right) \right].  
\end{eqnarray} 

In order to determine its minimum, $V_{\text{CW}}$ has to satisfy
\begin{eqnarray}
\frac{dV_{\text{CW}}}{dB_c^\mu} &=& 0,~~\mathrm{for~some~}B_c^\mu=b_\mu~,\label{gap1}\\
M_B^2 &=& \frac{d^2V_{\text{CW}}}{dB_c^\mu d{B_{c}}_{\mu}}\Big{|}_{B_c^\mu=b^\mu}>0.\label{mass1}
\end{eqnarray}

These conditions are met for a nontrivial $b^\mu$ when $b^2=v^2~\mathrm{exp}\left[ 1-\frac{2\lambda}{\beta(\lambda)}\right]$. Opting for the renormalization scale to coincide with the minimum of the Coleman-Weinberg potential implies $\lambda=\beta(\lambda)/2$. This equation can be iteratively solved, yielding
\begin{eqnarray}\label{lambdaCW}
	\lambda &=&\frac{\beta_{(\lambda=0)}}{2}+\cdots \nonumber\\
	&=&
	\frac{\xi_1^2(13f_0^4-5f_0^2 f_2^2+325 f_2^4)}{9216\pi^2}
	+\frac{5\xi_2^2(f_0^4-f_0^2 f_2^2+f_2^4)}{64\pi^2}  \nonumber\\
	&&
	+\frac{\xi_1 \xi_2(8f_0^4-5f_0^2 f_2^2+30 f_2^4)}{384\pi^2}
	+\mathcal{O}(\xi_i^4).
\end{eqnarray}  

Substituting the value of $\lambda$ in Eq.\eqref{eqVCW}, we have
\begin{eqnarray}\label{eqVcw}
V_{\text{CW}}\approx \frac{\beta_{(\lambda=0)}}{2} B_c^4\left[\ln\left(\frac{B_c^2}{v^2}\right)-\frac{1}{2} \right].  
\end{eqnarray}
\noindent For this solution, the dynamically generated mass for the bumblebee field, as given by \eqref{mass1}, is $M_B^2=4v^2\beta_{(\lambda=0)}$. Additionally, if the gravitational couplings $f_0$, $f_2$, $\xi_1$, and $\xi_2$ are sufficiently small close to the vacuum, $f_{0,2}\sim 10^{-8}$ as discussed in Ref.\cite{Salvio:2014soa}, $\lambda$ can be approximately vanishing near the vacuum, resulting in an approximately flat spacetime.

\section{Considerations of Potential Physical Scenarios}

An important aspect of the model under consideration involves the non-minimal gravitational couplings to the vector field, expressed as: 
\begin{eqnarray} \mathcal{L}_{nm}=\sqrt{-g}\left[\xi_1(B^{\mu}B^{\nu}-\frac{1}{4}B^2g^{\mu\nu})R_{\mu\nu} + \xi_2B^2R\right]. 
\end{eqnarray}
	
We will begin by analyzing the beta functions for the non-minimal gravitational couplings $\xi_1$ and $\xi_2$, as detailed in Eqs. \eqref{betaxi1} and \eqref{betaxi2}. Notably, both beta functions exhibit nontrivial fixed points at $\xi_1^* = \frac{g_l + 5g_t}{g_l^2 + 4g_l g_t + 7g_t^2}$ and $\xi_2^* = \frac{3g_t - g_l}{12 \left(g_l^2 + 3g_t^2\right)}$, respectively. It is important to note that these nontrivial fixed points fall outside the regime of perturbative validity, as they scale as $\xi_i^* \sim \frac{1}{g_i}$ due to the perturbative nature of the coupling constants in the model. Furthermore, for perturbative couplings, $\xi_1$ is asymptotically free ($\beta(\xi_1) < 0$), while $\xi_2$ can be asymptotically free ($\beta(\xi_2) < 0$) if $g_l < 3g_t$; otherwise, $\beta(\xi_2)$ becomes positive.
	
From \eqref{eqVcw}, we observe that the minimum of the effective CW potential for the bumblebee field corresponds to a nontrivial VEV $\langle B^{\mu}\rangle = v$. This nontrivial VEV for $B^{\mu}$ results in the dynamical generation of a LV gravitational term, $s^{\mu\nu}R_{\mu\nu} = \xi_1(b^{\mu}b^{\nu} - \frac{1}{4}b^2 g^{\mu\nu})R_{\mu\nu}$, as well as an Einstein-Hilbert term, $\xi_2 b^2 R$. We can analyze these contributions in two LV regimes: one where $b^2$ is very small and another where $b^2 \sim M_P^2$.
	
Experimental constraints provide bounds on the traceless LV tensor $s^{\mu\nu}$, with $s^{\mu\nu} = \xi_1(b^{\mu}b^{\nu} - \frac{1}{4}b^2 g^{\mu\nu})$ being constrained to less than $10^{-11}$ GeV$^2$ \cite{Kostelecky:2008ts}. Given that the beta function for $\xi_1$ is negative, this LV coupling will diminish further at higher energies, thereby restoring Lorentz symmetry in the deep UV. If $s^{\mu\nu}$ remains small, it is possible that the LV scale $v$ is related to the smallness of $s^{\mu\nu}$. In such a scenario, the trace part operator $\xi_2 b^2 R$ can be seen as a minor correction to the Einstein-Hilbert term, with its dynamical generation attributable to other beyond the SM field. For instance, in the original Agravity proposal \cite{Salvio:2014soa}, the generation of the Planck scale is attributed to a scalar field beyond the SM.

Another significant aspect of the present model concerns the vector field $B^\mu$, specifically its mass. While $B^\mu$ is massless at tree level, it acquires a mass through spontaneous Lorentz symmetry violation (LSV), with $M_B^2 = 4v^2\beta_{(\lambda=0)}$. This feature is particularly noteworthy because the bumblebee field could serve as a potential dark matter (DM) candidate. Recent studies suggest that massless (dark) photons cannot be sufficiently concentrated to form an event horizon \cite{Alvarez-Dominguez:2024pub}, implying that if DM is described by a vector field, it must possess mass. For ultralight vector DM particles \cite{Nelson:2011sf}, where $10^{-28}$ eV $\ll M_B< 1$ eV, the Lorentz violation (LV) scale could range from $10^{-19}\mathrm{GeV} \ll v \lesssim 10^{9}\mathrm{GeV}$. In this scenario, the LV scale would be significantly below the Planck scale, which is compatible with a near-vanishing cosmological constant. 

In a more speculative scenario, the bumblebee field could acquire a mass on the order of $M_B \sim 10~\mathrm{GeV}$. There are compelling empirical arguments suggesting that DM may have a mass around $10~\mathrm{GeV}$ \cite{Hooper:2012ft}. Assuming $\xi_1 \ll \xi_2 \sim 10^{-1}$ and $f_0 \sim f_2 \sim 10^{-8}$, the LV scale would be approximately $v \sim \frac{M_B}{\beta_{(\lambda=0)}} \sim 10^{19}~\mathrm{GeV} \sim M_P$. This estimation supports the concept of a large LV scale. In such a case, $\xi_1$ would need to be extremely small to ensure $s^{\mu\nu} < 10^{-11}~\mathrm{GeV}^2$. However, with $\xi_2$ being perturbative yet not exceedingly small, the term $\xi_2 b^2 R = \xi_2 v^2 R$ can be associated with the Einstein-Hilbert term $\frac{2}{\kappa^2}R$, where $\kappa^2$ can be interpreted as an effective gravitational constant.

In this scenario, for $g_l > 3g_t$, the running of $\xi_2$ is determined by a positive beta function, $\beta(\xi_2) \approx \frac{\lambda (g_l - 3g_t)}{12 \pi^2} > 0$. Since the effective gravitational coupling $\kappa$ scales as $1/\sqrt{\xi_2}$, this implies that $\kappa$ would be asymptotically free. Conversely, if $g_l < 3g_t$, $\beta(\xi_2)$ becomes negative, suggesting that the effective $\kappa$ would increase with the energy scale.

In these latter cases, however, the effective potential at its minimum is $V_{\text{min}} \sim -\beta_{(\lambda=0)} M_P^4$, resulting in a non-negligible negative cosmological constant, even considering $f_{0,2} \sim 10^{-8}$ as discussed in Ref. \cite{Salvio:2014soa}. This situation would require significant fine-tuning between contributions from the bumblebee field and other beyond-Standard Model fields to achieve a cosmological constant of an acceptable magnitude. While this scenario is intriguing, it may need further refinement, but it could serve as a basis for exploring asymptotically free (or safe) gravitational interactions. We plan to revisit this issue in future work.

\section{Final Remarks}\label{summary}

We formulated the agravity-bumblebee model. The importance of our study 
consists in the fact that, first, it can serve as a prototype for studying perturbative effects in LV gravity models, second, it is free of notorious difficulties of quantum gravity, that is, non-renormalizability for the absence of higher-derivative terms  and ghosts, in presence of such terms. Actually, one could expect this theory to be a fundamental one while the Einstein-Hilbert term arises as a quantum correction.

Our study is based on calculating the renormalization group functions. We use the methodology of renormalization group improvement \cite{Lehum:2019msl} to obtain the CW effective potential, and arrive at the dynamical generation of mass for the bumblebee field. One of the consequences of our studies consists in a possibility of a relation between the Planck mass and the Lorentz-breaking scale which, in principle, could indicate a fundamental nature for the Lorentz symmetry breaking.

Further continuation of our study could consists in its generalization for other gravity models, in particular, non-Riemannian ones, especially, metric-affine ones. We expect to pursue these aims in our next papers.

\acknowledgments
The work of A. Yu. P. has been partially supported by the CNPq project No.303777/2023-0. The work of A. C. L. has been partially supported by the CNPq project No. 404310/2023-0. P. J. P. would like to acknowledge the Brazilian agency CNPq, grant No. 307628/2022-1. The authors would like to thank Professor V. Alan Kostelecky for his useful comments.

\end{document}